\documentclass[10pt]{extarticle}
\usepackage[utf8]{inputenc}

\usepackage{authblk}
\usepackage{blindtext}
\usepackage[margin=0.5in]{geometry}
\usepackage[super]{natbib}
\usepackage{amssymb}
\usepackage{graphicx}
\usepackage{pifont}
\newcommand{\xmark}{\ding{55}}%
\usepackage{multirow}
\title{Reliability of two Embedded Atom Models for the Description of Au-Ag Nanoalloys}
\author[1,*]{M. Bon}
\author[1]{N. Ahmad}%
\author[1]{R. Erni}%
\author[2,*]{D. Passerone}

\affil[1]{\small Electron Microscopy Center, Empa, \"Uberlandstrasse 129, CH-8600, D\"ubendorf.\small}
\affil[2]{nanotech@surfaces, Empa, \"Uberlandstrasse 129, CH-8600, D\"ubendorf.\small}
\affil[*]{Correponding Authors: marta.bon@empa.ch\\ daniele.passerone@empa.ch\small}

\date{}

\begin{document}
\maketitle
  
\date{\today}

\begin{abstract}
The validation of embedded atom models (EAM) for modelling nanoalloys requires to verify both a faithful description of the individual phases and a convincing scheme for the mixed interactions. In this work, we present a systematic benchmarking of two widely adopted EAM parameterizations, i.e. by Foiles [\textit{S. M. Foiles et al. Phys. Rev. B
33, 7983 (1986)}] and by Zhou [\textit{X. W. Zhou et al. Phys. Rev. B, 69, 144113 (2004)}] with density functional theory calculations for the description of processes at Ag@Au nanoalloys surfaces and nanoclusters.\\
%
\end{abstract}

\maketitle

\section{\label{sec:Introduction}Introduction}
Bimetallic nanoparticles (NPs) exhibit unique optical, electronic, magnetic, and catalytic properties. Within this class of systems,  Ag@Au are of particular interest because they combine the excellent plasmonic response of Ag (with consequent interesting applications in Raman spectroscopy\cite{yang_galvanic_2014}) to the Au chemical stability\cite{krishnan_seed-mediated_2018} suppressing the oxidation of the Ag otherwise responsible for the NP morphology degradation. Currently, much experimental effort is devoted to the synthesis and morphology/shape control of these nanostructures and to the understanding of the complex evolution pathways to the desired nanosystem starting from the very early building blocks, i.e. clusters and NPs.\cite{qian_cluster-assembled_2010} In this context, a theoretical treatment is often successfully combined to experiments for systematically investigating thermodynamical properties and chemical ordering (alloying, mixing patterns, surface segregation).\cite{ferrando_nanoalloys:_2008} For this purpose, different approaches have been used, spanning from density functional theory (DFT) to classical atomistic simulations. The latter become necessary when dealing with large systems, e.g. clusters of hundreds of atoms, and modelling large time-scale processes (of the order of microseconds), such as deposition and growth, commonly studied through classical molecular dynamics (MD) or Monte Carlo (MC) simulations (see e.g. refs. \citenum{baletto_growth_2003} and \citenum{ muller_lattice_2005}).  In a classical description, electrons do not appear explicitly, but their effect is included in the parametrization of the force fields describing the interactions between atoms. A widely used class of atomistic potentials for transition and  noble metals is obtained with the embedded atom method (EAM),\cite{daw_embedded-atom_1984, daw_embedded-atom_1993} successfully applied to several metals at the nano-scale (see e.g. refs. \citenum{wang_monte_2004} and \citenum{matsumoto_estimation_2009}).
The main idea of the EAM formalism is that each $i$-th atom of the system is embedded in the electron density $\rho_i$ of the neighbouring atoms, and is subject to a potential $V_i$:
\begin{equation}
    V_i=\frac{1}{2}\sum_j \Phi_{ij}(r_{ij}) + F[\rho_i],
\end{equation}
where $\Phi(r_{ij})$ is a pairwise interaction, and $F[\rho_i]$ is the embedding functional, whose analytical form depends on the model used. At the very small scale (and even more for clusters of less than a hundred of atoms) size effects can become important and local coordination environments energetically unimportant in the bulk may play a role. A reparametrization based on a test set enlarged with cluster-specific geometries has for example proved to be effective in the case of aluminum nanosystems, where modified functional forms were also tested.\cite{jasper_2004}
When estimating whether the scope of application of existing EAM schemes can be extended to originally not considered physical situations (in this paper, e. g. diffusion and adsorption on small cluster surfaces) a validation is thus always required.\\
In this work, we focused on the assessment of two EAM potentials for a faithful Ag@Au NP description, without any reparametrization with respect to the bulk situations. 
We tested two commonly used EAM parametrizations in the nanoalloy classical simulations, i.e. the one by Foiles (EAM-Foiles)\cite{foiles_embedded-atom-method_1986} and by Zhou (EAM-Zhou),\cite{zhou_misfit-energy-increasing_2004} comparing these classical estimations to DFT calculations at PBE\cite{perdew_generalized_1996} and PBE+Grimme dispersion corrections\cite{grimme_consistent_2010} (PBE-D3) levels. This validation is based on the benchmarking of some selected properties key-players in the alloy nano-growth, faceting and segregation processes, i.e. lattice parameters, surface and adsorption energies, and diffusion barriers.\cite{xia_role_2013} At first, calculations were carried out on three different low-indexes Ag and Au surfaces, and secondly on a truncated Ag cubooctahedral cluster and on a triangular Ag nanoplate. In this way, we assessed whether a faithful description by the two EAMs - whose parametrization often comes from bulk calculations - is recovered even when periodicity breaks down and small-size effects become important. Based on our findings, we conclude that the two potentials considered here can be safely adopted for nanoscience applications.

\section{\label{sec:Methods}Methods}

\subsection{\label{subsec:SurfaceEnergies}Surface Energies}
For three different low-indexes Ag and Au surfaces, we calculated the surface energies, following the work of Fiorentini and Methfessel.\cite{fiorentini_extracting_1996} We here summarize the derivation of variant 4 discussed in ref.\citenum{fiorentini_extracting_1996}, which was proven to be the best among the four methods tested in that work. At convergence (with slabs made of \textit{enough} layers), the following relation stands:
\begin{equation}
    E^N_{slab} \approx 2\sigma + N E_{bulk},
\end{equation}
where $E^N_{slab}$ is the energy of the $N-$layered slab, $\sigma$ the surface energy, and energy of the bulk ($E_{bulk}$) can be calculated as the slope of the straight line fitting to all the slab total-energy data versus $N$.\cite{fiorentini_extracting_1996} Note that $E^N_{slab}$, $\sigma$, and $E_{bulk}$ are extensive quantities, and consequentially the final $\sigma$ must be divided by the number of atoms constituting a single layer. To respect the symmetry determined by the stacking sequence in the $z$ direction, for the \{100\} and \{110\} terminations we used four different slabs corresponding to 5, 7, 9 and 11 layers. For the \{111\} termination instead we constructed 6, 9, 12 and 15 layered-slabs. Following the work of Singh-Miller and Marzari,\cite{singh-miller_surface_2009} the \{100\} was sampled with a 16x16 k-mesh in the $xy$ plane, and the other terminations were simulated with the closest k-mesh corresponding to a supercell of similar size in real space (11x16 for \{110\} and 8x9 for the 2x2 \{111\} slab). These calculations were run using the pseudo- potential plane-wave QUANTUM-ESPRESSO code.\cite{Giannozzi_2009} Interactions with frozen cores were described by the norm-conservinng Vanderbilt pseudopotentials approximation,\cite{Hamann_optimized_2013} the kinetic-energy cutoff for the plane-wave basis are 150 Ryd for the wave function and 600 Ryd for the charge density respectively.  Results are reported in Table \ref{tab:LatticeParametersSurfaceEnergies}.

\subsection{\label{subsec:MetAdsDiff} Adsorption and Diffusion}
\begin{figure}[htp]
\centering
\includegraphics[]{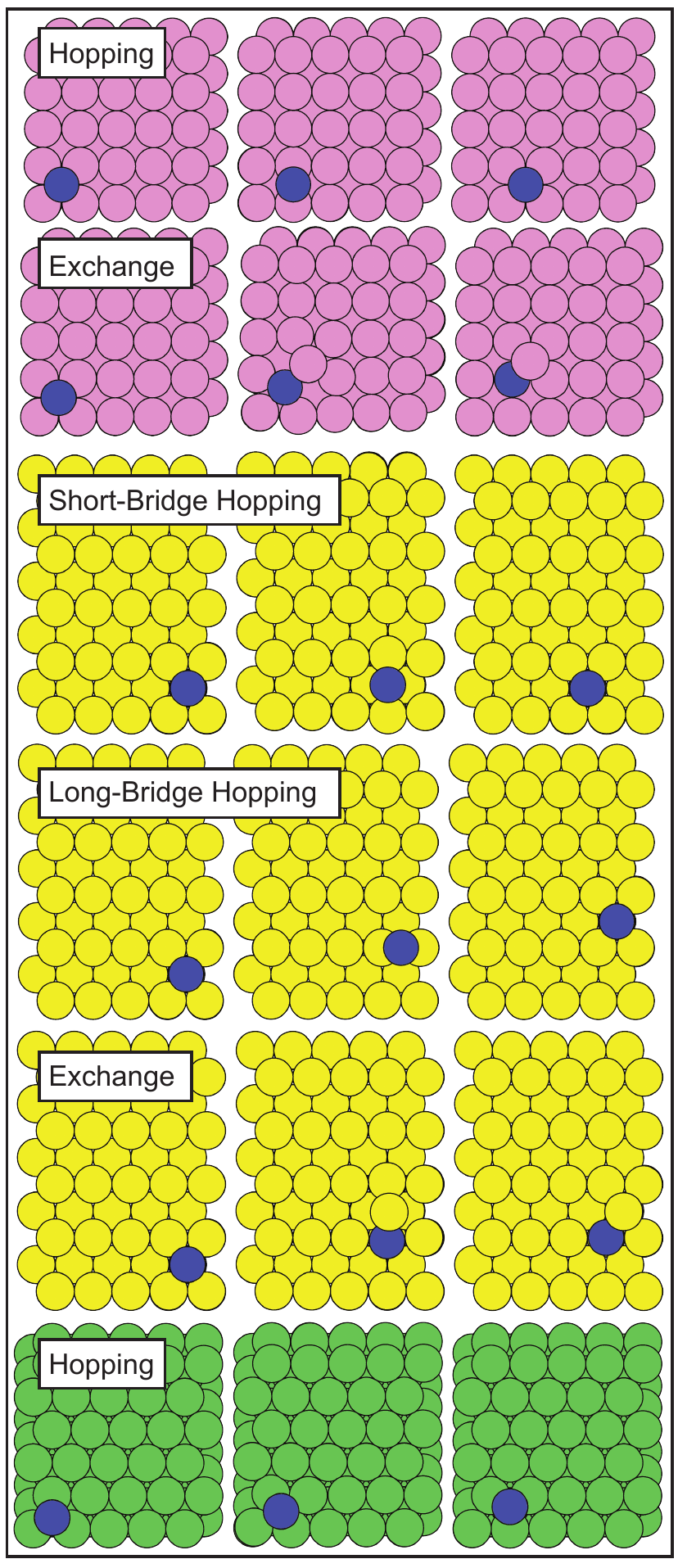}
\caption{\label{fig:SurfacesDiffusion} Diffusion mechanisms on the three simulated surfaces. The adatom is colored in blue and the surfaces are colored as explained in the following. Pink: \{100\}; Yellow: \{110\}; Green: \{111\}. In this study, surfaces and adatoms are of Ag or Au particles.}
\end{figure}

\begin{figure}[h]
\includegraphics[]{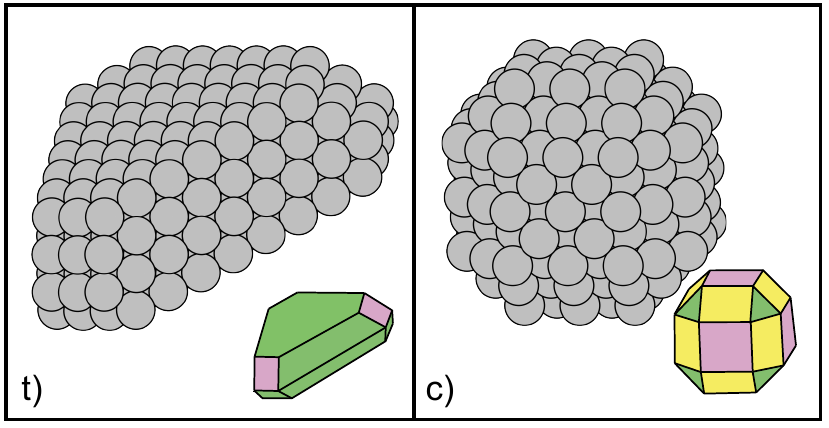}
\caption{\label{fig:ClustersShapes} The two simulated clusters shapes. \textbf{(t)} is the triangular nanoplate, and \textbf{(c)} the truncated cubooctahedron, assumed as cubic seed. On the right a schematic representation of the geometry. Green color indicates \{111\} surfaces, pink \{100\} and yellow \{110\} facets.}
\end{figure}

Adsorption sites and diffusion mechanisms on surfaces are shown in Fig. \ref{fig:SurfacesDiffusion} and the corresponding results are reported in Tables \ref{tab:AdsDiffAg100}-\ref{tab:AdsDiffAu111}.
The adsorption energies ($\mathrm{E}$) were calculated as:
\begin{equation}
    \label{eq:ads}
    \mathrm{E}=E_{tot}(0)+E_{at}-E_{tot}(1)
\end{equation}
where $E_{tot}(0)$ is the energy of the slab without the adatom, $E_{at}$ is the energy (equal to zero in the case of EAM) of the isolated atom and $E_{tot}(1)$ is the energy of the slab with the atom. Diffusion barriers ($E^*$) were evaluated through nudged elastic band (NEB) calculations,\cite{henkelman_climbing_2000} as implemented in the ASE python library,\cite{larsen_atomic_2017} which was interfaced with LAMMPS\cite{plimpton_fast_1995} and CP2K\cite{hutter_cp2k:_2014} calculators. For each process we used seven replicas linearly interpolated along the diffusion paths, a spring constant of 0.1 eV \AA$^{-1}$ and a convergence criterion for the forces equal to 0.005 eV \AA$^{-1}$.\\
At first, we considered a 5x5x6 supercell for \{100\} and \{110\} Ag and Au slabs and 5x6x5 for \{111\} surfaces. The bottom layer of each slab was considered as bulk layer and kept fixed during all the simulations.\\
Secondly, the same quantities were evaluated for the Ag clusters shown in Fig. \ref{fig:ClustersShapes}, i.e. a truncated cubooctahedron of 249 atoms and a triangular nanoplate of 295 atoms, exposing \{100\}, \{110\} and \{111\} facets. For these two systems results are reported in Tables \ref{tab:CUBICadatomAg}-\ref{tab:TRIANGULARadatomAu}. The motivation for choosing these particular model systems comes
from experiment where similar structures have been observed to undergo different growth pathways,\cite{yang_galvanic_2014, krishnan_seed-mediated_2018} stemming from differences in surface morphologies and relative ratios of the exposed planar surfaces.\\
We treated the valence electrons explicitly, using the DZVP basis set,\cite{vandevondele_gaussian_2007} whereas interactions with frozen atom cores were described with GTH pseudopotentials.\cite{krack_pseudopotentials_2005} We set the charge density plane wave cutoff equal to 400 Ry, and adopted the PBE exchange-correlation density functional.\cite{perdew_generalized_1996}  Note that the use of a finite basis set required to correct the adsorption energies in Eq.\ref{eq:ads} for the basis set superposition error (BSSE) by using the counterpoise method proposed by Boys and Bernardi. \cite{boys_calculation_1970}
We analyzed the effects of the inclusion of Grimme D3 dispersion corrections\cite{grimme_consistent_2010} on the calculated quantities which are known to bring a different contribution to cohesive energies (+14\% for Ag and +17\% for Au).\cite{rehr_van_1975} The choice of choosing PBE+D3 approximation was motivated by a recent study by Hoppe and M\"uller\cite{hoppe_first_2017} who performed a systematic comparison between different types of exchange and correlation functionals for the study of the Ag-Au (111) surfaces, with a particular focus on parameters determining the alloying and the segregation, and indicated PBE-D3 as the closest to the experimental evidence.
\section{Results}
\subsection{\label{subsec:ResLatticeSurfEn}Lattice Parameter and Surface Energies}
As clear from Table \ref{tab:LatticeParametersSurfaceEnergies} and in agreement with previous calculations,\cite{hoppe_first_2017, reckien_implementation_2012, chiter_effect_2016} the PBE functional tends to overestimate the lattice parameters, and the inclusion of Grimme corrections (PBE-D3) increases\cite{hoppe_first_2017} the values of the surface energies, consequently improving the agreement with the experiment.\\ Overall, the two EAMs qualitatively agree with DFT-PBE calculations, and predict equal lattice parameters for the Ag and Au and larger surface energies for Au (more evident in the case of the EAM-Zhou), in line with the experimental evidence. We point out that, similarly to the procedure adopted in ref. \citenum{singh-miller_surface_2009}, we do not consider Au surface reconstructions (which would be relevant on perfect high symmetry surfaces), also given the limited applicability of the surface reconstruction concepts known from bulk studies (i.e., herringbone for \{111\}, hexagonal for \{100\}, missing-row for \{110\}) to finite-size clusters.
\begin{table}
  \centering
  \caption{Lattice parameter (\AA), surface energies per atom $\sigma$ (eV/atom) and surface energies $\gamma$ (J/m$^{-2}$) for Ag and Au \{100\}, \{110\} and \{111\} surfaces. For the hetero-adsorption diffusion, $\mathrm{E}$ is the adsorption energy on the hollow site, and $\mathrm{E_{exc}}$ is the one associated to the final position after the exchange process (cfr. Fig. \ref{fig:SurfacesDiffusion}).}
  \label{tab:LatticeParametersSurfaceEnergies}
  \begin{tabular}{rcccccc}
    \hline
    & \multicolumn{1}{c}{} & \multicolumn{1}{c}{PBE} & \multicolumn{1}{c}{PBE+D3} & \multicolumn{1}{c}{EAM-Foiles} & \multicolumn{1}{c}{EAM-Zhou} &  Exp \\
    \hline
    \multicolumn{1}{l}{Ag} & a     & 4.15  & 4.09  & 4.09  & 4.09  & 4.09   \\
          & $\sigma_{100}$  & 0.43  & 0.64 & 0.37  & 0.51  &  \\
          & $\sigma_{110}$  & 0.66  & 1.09 & 0.56  & 0.80  &  \\
          & $\sigma_{111}$  & 0.35  & 0.55 & 0.28  & 0.41  &  \\
          & $\gamma_{100}$  & 0.81  & 1.24 & 0.70  & 0.98  &  \\
          & $\gamma_{110}$  & 0.62  & 1.05 & 0.54  & 0.77  &  \\
          & $\gamma_{111}$  & 0.66  & 1.05 & 0.54  & 0.79  & 1.25\footnote{\label{note1}Refs. [\citenum{boer_cohesion_1988,c._kittel_introduction_2004} ]}  \\
    \hline
    \multicolumn{1}{l}{Au} & a     & 4.15  & 4.11  & 4.08  & 4.08  & 4.08 \\
          & $\sigma_{100}$  & 0.47  & 0.73 & 0.48  & 0.53  &  \\
          & $\sigma_{110}$  & 0.71  & 1.10 & 0.72  & 0.82  &  \\
          & $\sigma_{111}$  & 0.34  & 0.58 & 0.35  & 0.41  &  \\
          & $\gamma_{100}$  & 0.87  & 1.40 & 0.92  & 1.02  & \\
          & $\gamma_{110}$  & 0.66  & 1.04 & 0.69  & 0.79  & \\
          & $\gamma_{111}$  & 0.63  & 1.10 & 0.67  & 0.79  & 1.50\textsuperscript{a}\\ 
    \hline
    \end{tabular}%
\end{table}%
\subsection{\label{subsec:ResAdsDiffSurf}Adsorption and Diffusion on Surfaces}
\subsubsection{On \{100\} Surfaces}
It is well known\cite{yu_ab_1997, elkoraychy_numerical_2015} that the Ag and Au adsorption process on clean \{100\} surfaces can occur only on hollow sites, and that the diffusion can take place through hopping or exchange (Fig. \ref{fig:SurfacesDiffusion}). As shown in Tables \ref{tab:AdsDiffAg100}-\ref{tab:AdsDiffAu100}, on both Ag and Au surfaces the adsorption energies of Au adatoms are higher compared to the ones of Ag and the adsorption process of the two adatoms is more favorable on the Au slab compared to the one on Ag\{100\}. Note that, when describing the hetero-adsorption diffusion, we calculated two adsorption energies, i.e. the one corresponding to the initial hollow site ($\mathrm{E}$) and the one associated to the in-channel position ($\mathrm{E_{exc}}$), final site of the exchange process (cfr. Fig. \ref{fig:SurfacesDiffusion}). One can note that for both adatoms, $\mathrm{E_{exc} > E}$. Together with the fact that the barriers ($\mathrm{E^*_{exc}}$) associated to the in-channel hetero Ag/Au and Au/Ag diffusion are lower than the corresponding activation energies for the hopping mechanisms ($\mathrm{ E^*_{hop}}$), this evidence indicates a strong tendency for the two species to intermix / alloy. On both surfaces, we observe that the EAM-Zhou strongly favors (compared to DFT and EAM-Foiles) the exchange among the hetero diffusion mechanisms. In fact, while EAM-Foiles and DFT calculations predict an absolute difference of at most 0.08 eV between the barriers associated to the possible processes ($\mathrm{E^*_{exc}}$ and $\mathrm{E^*_{hop}}$), the one calculated using EAM-Zhou is of 0.24 eV for Au/Ag and 0.38 eV for Ag/Au, in favor of the exchange process. Due to this prediction and because the exchange in-channel mechanism represents the first step for the adatom penetration in the slab, we expect that the EAM-Zhou favors the alloying more, compared to the other levels of theory.\\
Even in the case of Au self-diffusion, the exchange is favored to the hopping, while this trend is inverted for the self-diffusion on Ag\{100\}.\\
The two EAMs agree well with the energy trends predicted by DFT, with the unique significant deviation represented by the EAM-Zhou for which $\mathrm{ E^*_{hop}(Au/Au)< E^*_{hop}(Ag/Au)}$, contrary to what predicted by PBE, PBE+D3 and EAM-Foiles.

\begin{table}
  \centering
  \caption{Adsorption energies and diffusion barriers for Ag and Au adatoms on Ag \{100\} surface. For the hetero-adsorption diffusion, $\mathrm{E}$ is the adsorption energy on the hollow site, and $\mathrm{E_{exc}}$ is the one associated to the final position after the exchange process (cfr. Fig. \ref{fig:SurfacesDiffusion}). Units are in eV. }
  \label{tab:AdsDiffAg100}
    \begin{tabular}{cccccc}
    \hline
    \multicolumn{1}{c}{Adatom} & \multicolumn{1}{c}{} & \multicolumn{1}{c}{PBE} & \multicolumn{1}{c}{PBE+D3} & \multicolumn{1}{c}{EAM-Foiles} & \multicolumn{1}{c}{EAM-Zou}\\
    \hline
    \multirow{3}[2]{*}{Ag} & $\mathrm{E}$  & 2.37  & 2.77  & 2.39  & 2.19\\
          & $\mathrm{E^*_{hop}}$ & 0.46  & 0.40   & 0.48  & 0.50 \\
          & $\mathrm{E^*_{exc}}$ & 0.52  & 0.54  & 0.76  & 0.66\\
    \hline
    \multirow{4}[2]{*}{Au} & $\mathrm{E}$  & 2.99 & 3.32  & 3.27  & 2.79\\
          & $\mathrm{E_{exc}}$ & 3.02  & 3.43  & 3.51  & 3.22 \\
          & $\mathrm{ E^*_{hop}}$ & 0.55  & 0.50   & 0.65  & 0.71 \\
          & $\mathrm{E^*_{exc}}$ & 0.43  & 0.49  & 0.67  & 0.47\\
    \hline
    \end{tabular}%
\end{table}%
\begin{table}[h]
  \centering
  \caption{Adsorption energies and diffusion barriers for Ag and Au adatoms on Au \{100\} surface. Units are in eV.}
  \label{tab:AdsDiffAu100}
    \begin{tabular}{cccccc}
    \hline
    \multicolumn{1}{c}{Adatom} & \multicolumn{1}{c}{} & \multicolumn{1}{c}{PBE} & \multicolumn{1}{c}{PBE+D3} & \multicolumn{1}{c}{EAM-Foiles} & \multicolumn{1}{c}{EAM-Zhou}\\
    \hline
    \multirow{3}[1]{*}{Au} & $\mathrm{E}$  & 3.06  & 3.60  & 3.49  & 3.35\\
          & $\mathrm{E^*_{hop}}$ & 0.61  & 0.59  & 0.67  & 0.78 \\
          & $\mathrm{E^*_{exc}}$ & 0.11  & 0.09  & 0.35  & 0.40 \\
    \hline
    \multirow{4}[1]{*}{Ag} & $\mathrm{E}$    & 2.70  & 3.12  & 2.70   & 2.49 \\
          & $\mathrm{E_{exc}}$  & 2.88  & 3.26  & 2.55  & 2.25 \\
          & $\mathrm{E^*_{hop}}$ & 0.53  & 0.47  & 0.57  & 0.98 \\
          & $\mathrm{E^*_{exc}}$ & 0.19  & 0.16  & 0.50  & 0.60\\
    \hline
    \end{tabular}%

\end{table}%

\subsubsection{\label{subsubsec:RESAgSurf}On \{110\} Surfaces}
Likewise the case of \{100\} surfaces, on the \{110\} slabs the Ag and Au adatoms are located on hollow sites. On this termination, three possible diffusion mechanisms can take place. The first two are hopping processes, with the transition state located on long (LB-hopping) or short bridges (SB-hopping), while the third is a cross-channel diffusion via atomic exchange of the adsorbate with one atom of the channel wall (Fig. \ref{fig:SurfacesDiffusion}).\cite{perkins_influence_1995, elkoraychy_numerical_2015}\\
As shown in Tables \ref{tab:AdsDiffAg110} and \ref{tab:AdsDiffAu110}, the Ag adsorption is disfavored compared to the one of the Au adatom for both surfaces at all levels of theory. On the contrary, for both Ag and Au \{110\} surfaces  the adsorption energies after the occurrence of the in-channel hetero adsorption ($\mathrm{E_{exc}}$) are higher compared to the ones of the hollow site in the DFT description. An opposite trend is instead recovered by applying the EAM approximation.\\
The SB-hopping is the most favored diffusion mechanism for Ag/Ag and Ag/Au, with a unique exception represented by the PBE+D3 level of theory. The addition of dispersion corrections leads the Ag/Ag exchange to be the mechanism associated to the lowest barrier among the possible self-diffusion processes. Further disagreement between the different four predictions is found in the description of the Ag/Au diffusion. While for PBE and EAM-Foiles $\mathrm{E^*_{SB}<E^*_{exc}}$, the PBE+D3 and the EAM-Zhou predict the opposite result. Last, the SB-hopping is the most favorable mechanism for Au/Au self-diffusion at all the tested levels of theory. 

\begin{table}[h]

  \centering
  \caption{Adsorption energies and diffusion barriers for Ag and Au adatoms on Ag \{110\} surface. Units are in eV. }
  \label{tab:AdsDiffAg110}
    \begin{tabular}{cccccc}
    \hline
        \multicolumn{1}{c}{Adatom} & \multicolumn{1}{c}{} & \multicolumn{1}{c}{PBE} & \multicolumn{1}{c}{PBE+D3} & \multicolumn{1}{c}{EAM-Foiles} & \multicolumn{1}{c}{EAM-Zhou}\\
        \hline
        \multirow{4}[2]{*}{Ag} & $\mathrm{E}$  & 2.47  & 2.84  & 2.65  & 2.56\\
              & $\mathrm{E^*_{LB}}$ & 0.73  & 0.77  & 0.83  & 1.11 \\
              & $\mathrm{E^*_{SB}}$ & 0.36  & 0.30   & 0.32  & 0.25 \\
              & $\mathrm{E^*_{exc}}$ & 0.37  & 0.26  & 0.42  & 0.29\\
        \hline
        \multirow{5}[2]{*}{Au} & $\mathrm{E}$  & 3.14  & 3.53  & 3.59  & 3.25\\
              & $\mathrm{E_{exc}}$ & 3.05  & 3.50  & 3.73  & 3.48 \\
              & $\mathrm{E^*_{LB}}$ & 0.77  & 0.83  & 1.13  & 1.44 \\
              & $\mathrm{E^*_{SB}}$ & 0.44  & 0.38  & 0.30   & 0.30 \\
              & $\mathrm{E^*_{exc}}$ & 0.41  & 0.28  & 0.36  & 0.25\\
     \hline
        \end{tabular}%

\end{table}%

\begin{table}[h]

\caption{Adsorption energies and diffusion barriers for Au and Ag adatoms on Au \{110\} surface. Units are in eV.}
\label{tab:AdsDiffAu110}
  \centering
    \begin{tabular}{cccccc}
    \hline
    \multicolumn{1}{c}{Adatom} & \multicolumn{1}{c}{} & \multicolumn{1}{c}{PBE} & \multicolumn{1}{c}{PBE+D3} & \multicolumn{1}{c}{EAM-Foiles} & \multicolumn{1}{c}{EAM-Zhou}\\
    \hline
    \multirow{4}[2]{*}{Au} & $\mathrm{E}$  & 2.93  & 2.59  & 3.71  & 3.65 \\
          & $\mathrm{E^*_{LB}}$ & 0.66  & 0.72  & 1.04  & 1.26 \\
          & $\mathrm{E^*_{SB}}$ & 0.45  & 0.36  & 0.25  & 0.29 \\
          & $\mathrm{E^*_{exc}}$ & 0.48  & 0.38  & 0.40   & 0.38 \\
    \hline
    \multirow{5}[2]{*}{Ag} & $\mathrm{E}$  & 2.53  & 2.99  & 2.91  & 2.78 \\
          & $\mathrm{E_{exc}}$ & 2.81  & 3.08  & 2.80   & 2.65 \\
          & $\mathrm{E^*_{LB}}$ & 0.69  & 0.76  & 1.13  & 1.45 \\
          & $\mathrm{E^*_{SB}}$ & 0.38  & 0.33  & 0.30  & 0.3 \\
          & $\mathrm{E^*_{exc}}$ & 0.44  & 0.31  & 0.70   & 0.24 \\
    \hline
    \end{tabular}%

\end{table}%

\begin{table}[h]

  \centering
  \caption{Adsorption energies and diffusion barriers for Ag and Au adatoms on Ag \{111\} surface. Units are in eV.}
  \label{tab:AdsDiffAg111}
    \begin{tabular}{cccccc}
    \hline
    \multicolumn{1}{c}{Adatom} & \multicolumn{1}{c}{} & \multicolumn{1}{c}{PBE} & \multicolumn{1}{c}{PBE+D3} & \multicolumn{1}{c}{EAM-Foiles} & \multicolumn{1}{c}{EAM-Zhou}\\
    \hline
    \multirow{3}[1]{*}{Ag} & $\mathrm{E_{fcc}}$ & 2.16  & 2.47  & 2.14  & 1.91 \\
          & $\mathrm{E_{hcp}}$  & 2.16  & 2.47  & 2.14  & 1.89 \\
          & $\mathrm{E^*_{hop}}$  & 0.06  & 0.06  & 0.06  & 0.05 \\
    \hline
    \multirow{3}[1]{*}{Au} & $\mathrm{E_{fcc}}$ & 2.68  & 3.12  & 2.92  & 2.37 \\
          & $\mathrm{E_{hcp}}$ & 2.68  & 3.11  & 2.92  & 2.35 \\
          & $\mathrm{E^*_{hop}}$ & 0.08  & 0.09  & 0.05  & 0.04\\
    \hline
    \end{tabular}%

\end{table}%
\begin{table}[htbp]

  \centering
  \caption{Adsorption energies and diffusion barriers for Ag and Au adatoms on Au \{111\} surface. Units are in eV.}
  \label{tab:AdsDiffAu111}
    \begin{tabular}{cccccc}
    \hline
    \multicolumn{1}{c}{Adatom} & \multicolumn{1}{c}{} & \multicolumn{1}{c}{PBE} & \multicolumn{1}{c}{PBE+D3} & \multicolumn{1}{c}{EAM-Foiles} & \multicolumn{1}{c}{EAM-Zhou}\\
    \hline
    \multirow{3}[1]{*}{Au} & $\mathrm{E_{fcc}}$ & 2.31  & 3.19  & 3.03  & 2.86\\
          & $\mathrm{E_{hcp}}$  & 2.30  & 3.19  & 3.03  & 2.85 \\
          & $\mathrm{E^*_{hop}}$ & 0.11  & 0.12   & 0.02  & 0.05 \\
    \hline
    \multirow{3}[1]{*}{Ag} & $\mathrm{E_{fcc}}$  & 2.07  & 2.51  & 2.35  & 2.20 \\
          & $\mathrm{E_{hcp}}$  & 2.07  & 2.45  & 2.35  & 2.18 \\
          & $\mathrm{E^*_{hop}}$ & 0.08  & 0.09  & 0.05  & 0.05 \\
    \hline
    \end{tabular}%

\end{table}%

\subsubsection{\label{subsubsec:RESAgSurf111}On \{111\} Surfaces}
The \{111\} surfaces exhibit two binding sites for Ag and Au adatoms, corresponding to fcc (fcc-hollow) and hcp (hcp-hollow) stacking and characterized by very similar adsorption energies, as shown in Tables \ref{tab:AdsDiffAg111}, \ref{tab:AdsDiffAu111}. Likewise the case of \{100\} and \{110\} surfaces, the adsorption energies for the Au adatom are always higher than the ones of Ag. The almost barrier-less diffusion process between the two occurs via hopping passing through the bridge position, as shown in Fig. \ref{fig:SurfacesDiffusion}.\\
The EAMs tested show an overall agreement with DFT calculations in the description of the diffusion processes.

\subsection{Final Remarks on Adsorption and Diffusion on Surfaces}
The two EAMs overall agree in the prediction of the surface properties, albeit showing some tiny but crucial differences between the two models in the description of the energy barriers of diffusion processes associated to intermixing phenomena (e.g. in-channel exchange). These differences can address simulations to different final chemical-ordering scenarios. In particular, we would like to stress that the EAM-Zhou shows lower energy barriers for exchange processes as compared to EAM-Foiles, and to the ab-initio estimations, and as a consequence will lead to a more expressed tendency to alloying and possible Ag surface segregation in the Ag@Au growth description. 
\begin{figure}[h]
\includegraphics[]{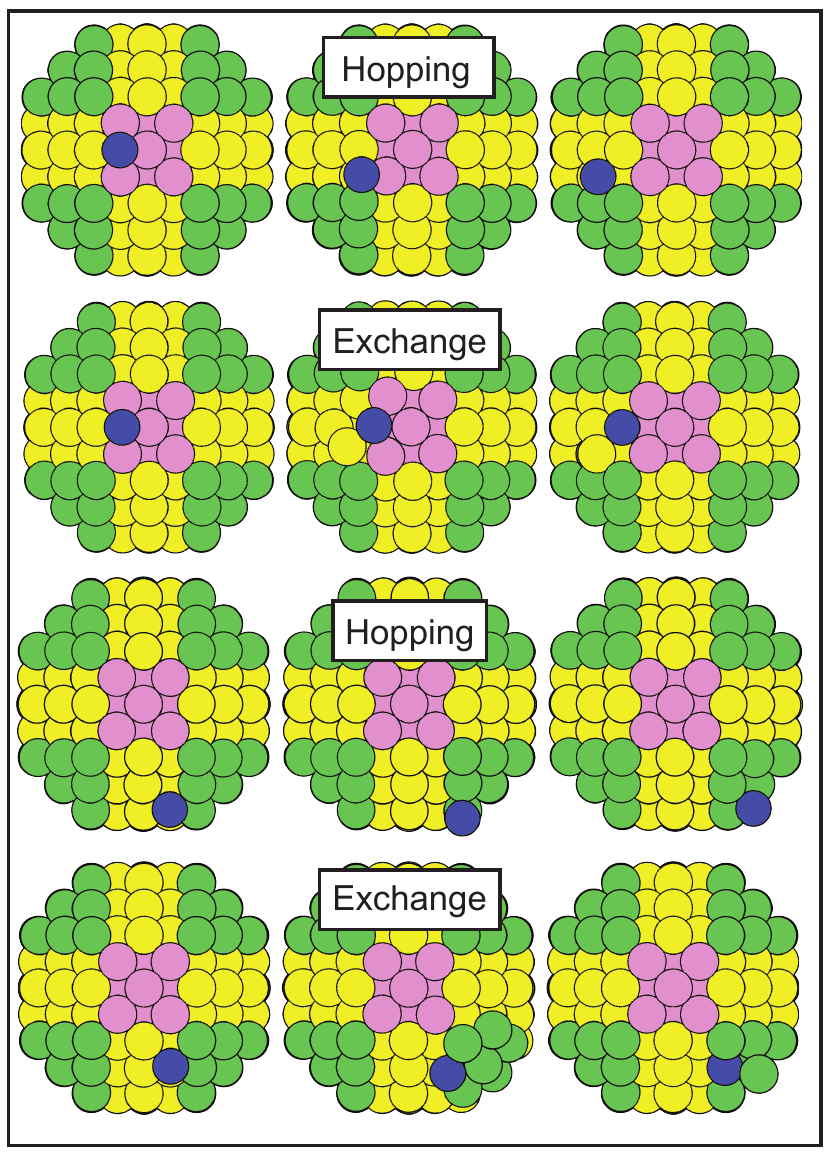}
\caption{\label{fig:CubicSurfacesDiffusion} Inter-facets diffusion processes on the cubooctahedral cluster (described in Fig. \ref{fig:ClustersShapes}). Same color scheme as in Fig. \ref{fig:SurfacesDiffusion}.}
\end{figure}
\begin{figure}[h]
\includegraphics[]{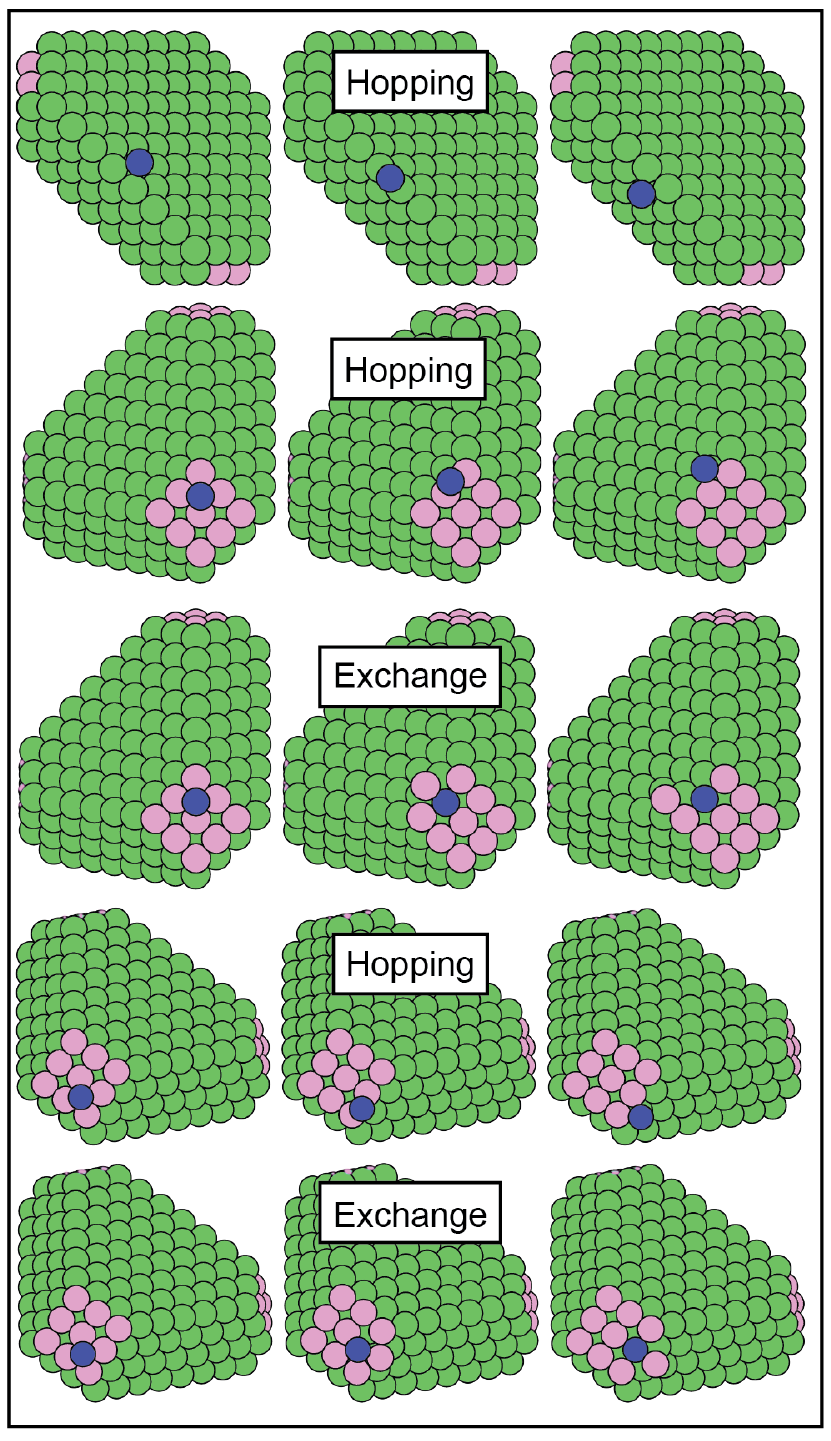}
\caption{\label{fig:TriangularSurfacesDiffusion} Inter-facets diffusion processes on the triangular cluster (described in Fig. \ref{fig:ClustersShapes}). Same color scheme as in Fig. \ref{fig:SurfacesDiffusion}.}
\end{figure}
\subsection{\label{subsec:ResAdsDiffClus}Adsorption and Diffusion on Clusters}
We now analyze the adsorption and self- and hetero- diffusion mechanisms on two clusters, i.e. a 3D-cubooctahedron and a triangular nanoplate, comparing the results obtained with the ones of the corresponding low-indexes 2D-periodic surfaces.
\subsubsection{\label{subsubsec:RESAgAdsCub}On the Ag Cubooctahedral Cluster}
As shown in Figs. \ref{fig:ClustersShapes} and \ref{fig:CubicSurfacesDiffusion}, the truncated cubooctahedron is made of \{100\} and \{110\} side facets with \{111\} corners.\\
The adsorption energies on these facets change at most around the 5\% with respect to the surface case and the major differences mostly concern PBE+D3 and EAM-Zhou. The most remarkable discrepancy from the infinite slab interests the description of the stability of the fcc-adsorption sites in the 6-atom corners \{111\}. Due to the small dimensions of this facet, this position is not predicted to be stable for the Ag adatom in DFT and EAM-Zhou descriptions, while configures as Au-adsorption site only at DFT-PBE and PBE-D3 levels.\\

The cluster can be subject to intra-facets diffusion processes such as hopping and exchange on the \{100\} and \{110\} surfaces and to hopping mechanisms on the \{111\} corners. On the \{100\} and \{110\} all the possible mechanisms are captured, with a unique exception represented by the EAM-Zhou that does not associate the LB-hopping to any minimum energy path for the Au diffusion. Instead, it favors an intermediate configuration that can be most likely associated to an exchange process. As a consequence of the already discussed instability of the hollow-fcc adsorption sites, on the \{111\} the hopping mechanism is not always recovered.\\
Overall, the barriers considerably change from the infinite surface case, and the largest deviations are recorded for the exchange and SB-hopping on the \{110\}. As expected, in general these differences are much more pronounced for DFT compared to the EAM approximation.\\
Additional inter-facets diffusion mechanisms, shown in Fig. \ref{fig:CubicSurfacesDiffusion}, are the hopping and exchange between \{110\} and \{111\} and between \{110\} and \{100\} facets. We point out a remarkable deviation of EAM-Zhou compared to the DFT trends concerning the description of the self-diffusion from \{100\} to \{110\} (Table \ref{tab:CUBICadatomAg}). Again, EAM-Zhou fails into the description of the hopping mechanism, which is instead recovered by EAM-Foiles and DFT calculations. Similar discrepancies are found in the representation of the Au inter-diffusion from \{100\} to \{110\} and from \{110\} to \{111\}, where both EAMs tend not to allow hopping mechanisms (Table \ref{tab:CUBICadatomAu}).\\
Among the inter-facets diffusion mechanisms, the most favored is the hopping from the \{111\} corners, associated to an energy barrier one order of magnitude lower compared to the others. This means that when undergoing growth processes, the \{111\} facets will expand, with progressive disappearance of \{110\} and \{100\} surfaces. 

\begin{table}[htbp]

  \centering
  \caption{Adsorption energies and diffusion barriers for Ag adatom on the cubooctahedral Ag seed. Units are in eV.}
  \label{tab:CUBICadatomAg}%
    \begin{tabular}{cccccc}
    \hline
          & & \multicolumn{1}{c}{PBE} & \multicolumn{1}{c}{PBE+D3} & \multicolumn{1}{c}{EAM-Foiles} & \multicolumn{1}{c}{EAM-Zhou} \\
    \hline
    \multicolumn{6}{c}{Intra-facet Diffusion}\\[5pt]
    \{100\} & $\mathrm{E}$  & 2.30  & 2.60  & 2.41  & 2.20\\
          & $\mathrm{E^*_{hop}}$ & 0.43  & 0.45  & 0.48  & 0.49 \\
          & $\mathrm{E^*_{exc}}$ & 0.45  & 0.56  & 0.78  & 0.73 \\[5pt]
    \{110\} & $\mathrm{E}$  & 2.49  & 2.82  & 2.68  & 2.62 \\
          & $\mathrm{E^*_{LB}}$ & 0.79  & 0.83  & 0.85  & 1.14 \\
          & $\mathrm{E^*_{SB}}$ & 0.24  & 0.17  & 0.32  & 0.24 \\
          & $\mathrm{E^*_{exc}}$ & 0.10  & 0.02  & 0.36  & 0.22 \\[5pt]
    \{111\} & $\mathrm{E_{fcc}}$ & \xmark & \xmark & 2.17  & \multicolumn{1}{c}{\xmark} \\
          & $\mathrm{E_{hcp}}$ & 2.03  & 2.36  & 2.20  & 1.98 \\
          & $\mathrm{E^*_{hop}}$ & \xmark & \xmark & 0.04  & \xmark\\[5pt]
    \hline
    \multicolumn{6}{c}{Inter-facets Diffusion}\\[5pt]
    \{100\} $\rightarrow$ \{110\} & $\mathrm{E^*_{hop}}$ & 0.26  & 0.28  & 0.49  & \multicolumn{1}{c}{\xmark}\\
          & $\mathrm{E^*_{exc}}$ & 0.26  & 0.27  & 0.36  & 0.32 \\[5pt]
    \{110\}  $\rightarrow$  \{100\} & $\mathrm{E^*_{hop}}$ & 0.45  & 0.52  & 0.77  & \multicolumn{1}{c}{\xmark}\\
          & $\mathrm{E^*_{exc}}$ & 0.45  & 0.50  & 0.36  & 0.32 \\[5pt]
   \{110\}  $\rightarrow$  \{111\} & $\mathrm{E^*_{hop}}$ & 0.86  & 1.00  & 0.84  & 0.66 \\
          & $\mathrm{E^*_{exc}}$ & 0.25  & 0.26  & 0.32  & 0.21 \\[5pt]
    \{111\}  $\rightarrow$  \{110\} & $\mathrm{E^*_{hop}}$ & 0.26  & 0.48  & 0.32  & 0.39 \\
    & $\mathrm{E^*_{exc}}$ & 0.09  & 0.02  & 0.08  & 0.02 \\
    \hline
    \end{tabular}%

\end{table}%

\begin{table}[h]

  \centering
  \caption{Adsorption energies and diffusion barriers for Au adatom on the cubooctahedral Ag seed. Units are in eV.}
  \label{tab:CUBICadatomAu}%
    \begin{tabular}{cccccc}
    \hline
    & & \multicolumn{1}{c}{PBE} & \multicolumn{1}{c}{PBE+D3} & \multicolumn{1}{c}{EAM-Foiles} & \multicolumn{1}{c}{EAM-Zhou} \\
    \hline
    \multicolumn{6}{c}{Intra-facet Diffusion} \\[5pt]
    \{100\} & $\mathrm{E}$  & 2.97  & 3.27  & 3.28  & 2.81 \\
          & $\mathrm{E_{exc}}$ & 2.94  & 3.29  & 3.54  & 3.27 \\
          & $\mathrm{E^*_{hop}}$ & 0.52  & 0.44  & 0.64  & 0.70 \\
          & $\mathrm{E^*_{exc}}$ & 0.39  & 0.46  & 0.58  & 0.43\\[5pt]
    \{110\} & $\mathrm{E}$  & 3.13  & 3.50  & 3.63  & 3.35 \\
          & $\mathrm{E_{exc}}$ & 3.12  & 3.50  & 3.78  & 3.58 \\
          & $\mathrm{E^*_{LB}}$ & 0.80  & 0.85  & 1.15  & \xmark \\
          & $\mathrm{E^*_{SB}}$ & 0.30  & 0.22  & 0.29  & 0.22 \\
          & $\mathrm{E^*_{exc}}$ & 0.08  & 0.29  & 0.30  & 0.17 \\[5pt]
    \{111\} & $\mathrm{E_{fcc}}$ & 2.63  & 2.88  & \xmark & \xmark \\
          & $\mathrm{E_{hcp}}$ & 2.68  & 3.01  & 3.00  & 2.52 \\
          & $\mathrm{E^*_{hop}}$ & 0.04  & 0.05  & \xmark & \xmark \\[5pt]
    \hline
    \multicolumn{6}{c}{Inter-facets Diffusion} \\[5pt]
    \{100\} $\rightarrow$ \{110\} & $\mathrm{E^*_{hop}}$ & 0.46  & 0.55  & \xmark & 0.14 \\
          & $\mathrm{E^*_{exc}}$ & 0.28  & 0.12  & 0.27  & 0.19 \\[5pt]
    \{110\} $\rightarrow$ \{100\} & $\mathrm{E^*_{hop}}$ & 0.62  & 0.77  & \xmark & 0.68 \\
          & $\mathrm{E^*_{exc}}$ & 0.43  & 0.32  & 0.72  & 0.88 \\[5pt]
    \{110\} $\rightarrow$ \{111\} & $\mathrm{E^*_{hop}}$ & 0.91  & 1.13  & \xmark & \xmark \\
          & $\mathrm{E^*_{exc}}$ & 0.61  & 0.57  & 0.43  & 0.34\\[5pt]
    \{111\} $\rightarrow$ \{110\} & $\mathrm{E^*_{hop}}$ & 0.22  & 0.50  & \xmark & \xmark  \\
    & $\mathrm{E^*_{exc}}$ & 0.08  & 0.01  & 0.11  & 0.08 \\
    \hline
    \end{tabular}%

\end{table}%

\begin{table}[h]

  \centering
  \caption{Adsorption energies and diffusion barriers for Ag adatom on the triangular Ag seed. Units are in eV.}
  \label{tab:TRIANGULARadatomAg}%
    \begin{tabular}{cccccc}
    \hline
          & \multicolumn{1}{c}{} & \multicolumn{1}{c}{PBE} & \multicolumn{1}{c}{PBE+D3} & \multicolumn{1}{c}{EAM-Foiles} & \multicolumn{1}{c}{EAM-Zhou}\\
    \hline
    \multicolumn{6}{c}{Intra-facet Diffusion}\\[5pt]
    \{111\} & $\mathrm{E_{fcc}}$ & 2.01  & 2.37 & 2.20  & 1.96 \\
          & $\mathrm{E_{hcp}}$ & 2.02  & 2.38  & 2.19  & 1.94 \\
          & $\mathrm{E^*_{hop}}$ & 0.04  & 0.05  & 0.05  & 0.05 \\[5pt]
    \{111\}L & $\mathrm{E_{fcc}}$ & 1.96  & 2.30  & 2.19  & 1.96 \\
          & $\mathrm{E_{hcp}}$ & 1.97  & 2.31  & 2.19  & 1.95 \\
          & $\mathrm{E^*_{hop}}$ & 0.04  & 0.05  & 0.05  & 0.03 \\[5pt]
    \{100\} & E  & 2.15  & 2.40  & 2.46  & 2.21 \\
          & $\mathrm{E^*_{hop}}$ & 0.40  & 0.32  & 0.47  & 0.17 \\
          & $\mathrm{E^*_{exc}}$ & 0.41  & 0.45  & 0.78  & 0.31 \\[5pt]
    \hline
    \multicolumn{6}{c}{Inter-facets Diffusion} \\[5pt]
    \{100\} $\rightarrow$ \{111\} & $\mathrm{E^*_{hop}}$ & 0.49  & 0.52  & 0.59  & 0.74 \\
          & $\mathrm{E^*_{exc}}$ & 0.44  & 0.44  & 0.37  & 0.29 \\[5pt]
    \{111\}  $\rightarrow$ \{100\} & $\mathrm{E^*_{hop}}$ & 0.32  & 0.37  & 0.34  & 0.52 \\
          & $\mathrm{E^*_{exc}}$ & 0.10  & 0.26  & 0.12  & 0.02 \\[5pt]
    \{100\} $\rightarrow$ \{111\}L & $\mathrm{E^*_{hop}}$ & 0.53  & 0.56  & 0.59  & 0.76 \\
          & $\mathrm{E^*_{exc}}$ & 0.24  & 0.47  & 0.64  & 0.23 \\[5pt]
    \{111\}L  $\rightarrow$ \{100\} & $\mathrm{E^*_{hop}}$ & 0.30  & 0.34  & 0.34  & 0.52 \\
          & $\mathrm{E^*_{exc}}$ & 0.03  & 0.32  & 0.39  & 0.02 \\[5pt]
    \{111\}  $\rightarrow$ \{111\}L & $\mathrm{E^*_{hop}}$ & 0.19  & 0.28  & 0.36  & 0.60 \\[5pt]
    \{111\}L  $\rightarrow$ \{111\} & $\mathrm{E^*_{hop}}$ & 0.18  & 0.28  & 0.36  & 0.59 \\[5pt]
    \hline
    \end{tabular}%

\end{table}%
\begin{table}[h]
  \centering
  \caption{Adsorption energies and diffusion barriers for Au adatom on the triangular Ag seed. Units are in eV.}
  \label{tab:TRIANGULARadatomAu}%
    \begin{tabular}{lccccc}
    \hline
          & \multicolumn{1}{c}{} & \multicolumn{1}{c}{PBE} & \multicolumn{1}{c}{PBE+D3} & \multicolumn{1}{c}{EAM-Foiles} & \multicolumn{1}{c}{EAM-Zhou} \\
    \hline
    \multicolumn{6}{c}{Intra-facet Diffusion}  \\[5pt]
    \{111\} & $\mathrm{E_{fcc}}$ & 2.52  & 2.93  & 2.98  & 2.41 \\
          & $\mathrm{E_{hcp}}$ & 2.52  & 2.93  & 2.98  & 2.38 \\
          & $\mathrm{E^*_{hop}}$ & 0.05  & 0.07  & 0.04  & 0.04 \\[5pt]
    \{111\}L & $\mathrm{E_{fcc}}$ & 2.53  & 2.92  & 2.98  & 2.41 \\
          & $\mathrm{E_{hcp}}$ & 2.54  & 2.92  & 2.98  & \multicolumn{1}{c}{\xmark} \\
          & $\mathrm{E^*_{hop}}$ & 0.05  & 0.07  & 0.04  & \multicolumn{1}{c}{\xmark} \\[5pt]
    \{100\} & $\mathrm{E}$  & 2.80  & 3.08  & 3.33  & 2.82 \\
          & $\mathrm{E^*_{exc}}$ & 2.74  & 3.06
          & 3.61  & 3.35 \\
          & $\mathrm{E^*_{hop}}$ & 0.47  & 0.39  & 0.60  & 0.17 \\
          & $\mathrm{E^*_{exc}}$ & 0.43  & 0.38  & 0.19  & 0.31 \\
    \hline
    \multicolumn{6}{c}{Inter-facets Diffusion}  \\[5pt]
    \{100\} $\rightarrow$ \{111\} & $\mathrm{E^*_{hop}}$ & 0.57  & 0.58  & 0.83  & 1.02 \\
          & $\mathrm{E^*_{exc}}$ & 0.58  & 0.50  & 0.22  & 0.31\\[5pt]
   \{111\}  $\rightarrow$ \{100\} & $\mathrm{E^*_{hop}}$ & 0.33  & 0.36  & 0.48  & 0.67 \\
          & $\mathrm{E^*_{exc}}$ & 0.37  & 0.33  & 0.16  & 0.30 \\[5pt]
    \{100\} $\rightarrow$ \{111\}L & $\mathrm{E^*_{hop}}$ & 0.57  & 0.58  & 0.82  & 1.02 \\
          & $\mathrm{E^*_{exc}}$ & 0.57  & 0.50  & 0.43  & 0.37 \\[5pt]
    \{111\}L  $\rightarrow$ \{100\} & $\mathrm{E^*_{hop}}$ & 0.29  & 0.35  & 0.54  & 0.72 \\
          & $\mathrm{E^*_{exc}}$ & 0.27  & 0.31  & 0.44  & 0.67 \\[5pt]
    \{111\}  $\rightarrow$ \{111\}L & $\mathrm{E^*_{hop}}$ & 0.16  & 0.30  & 0.53  & 0.75 \\[5pt]
    \{111\}L  $\rightarrow$ \{111\} & $\mathrm{E^*_{hop}}$ & 0.17  & 0.31  & 0.53  & 0.53 \\
    \hline
    \end{tabular}%

\end{table}%

\subsubsection{\label{subsubsec:RESAgAdsTr}On the Ag Triangular Cluster}
As shown in Figs. \ref{fig:ClustersShapes} and \ref{fig:TriangularSurfacesDiffusion}, the triangular seed mainly exposes \{111\} facets with \{100\} corners, and therefore is characterized by surface hopping on the \{111\} surfaces, and hopping and exchange on the \{100\}. Furthermore, inter-facets diffusion can involve the large and the lateral \{111\} surfaces \{111\} surfaces through hopping, and the \{100\} corners and the large \{111\} facets or the \{100\} corners and the lateral \{111\} facets  via hopping and exchange. The just mentioned processes are represented in Fig. \ref{fig:TriangularSurfacesDiffusion}, and the corresponding adsorption energies and diffusion barriers are reported in Tables \ref{tab:TRIANGULARadatomAg} and \ref{tab:TRIANGULARadatomAu} (note that the lateral \{111\} surfaces are labeled as \{111\}L). \\ 
The most favorable adsorption site on the cluster is the hollow position on the \{100\} corners for both Ag and Au adatoms. Overall, DFT estimates lower adsorption energies compared to the corresponding infinite slab with differences of at most 12\% (associated to the Ag adsorption on \{100\} surface), while EAMs show the opposite trend, with smaller differences (at most of the 3\%). We must point out that EAM-Zhou does not allow the Au adsorption on the hollow-fcc position of the \{111\} lateral facets, in contrast with DFT and EAM-Foiles. We strongly believe that this is an effect of the small size of the facet, as found in the description of the cubooctahedral \{111\} corners.\\
Likewise the case of the cubooctahedral seed, the barriers change from the ones describing diffusion on the corresponding infinite surface. The trends provided by the EAMs are the same of the ones from DFT for the intra-facet diffusion processes, albeit observing in the EAMs a more pronounced tendency towards hopping on \{100\} compared to the other levels of theory. For both adatoms,  the barriers describing hopping on \{111\} and \{111\}L facets do not present significant differences from the infinite surface at all levels of theory, indicating that the intra-facet kinetics on these surfaces is not largely affected by its dimensions.\\Inter-facets diffusion is overall well represented by the EAM parametrization. A few deviations from DFT trends are recovered in the EAM-Foiles description self-diffusion involving the \{111\} short facets and in a more pronounced tendency towards exchange (when possible) in the EAM-Zhou estimations.

\subsection{Final Remarks on Adsorption and Diffusion on Clusters}
Although observing some (expected) deviations from the DFT predictions, the two EAMs well represent the majority of the processes on both nano-systems considered in this study. For both clusters, the barriers describing the diffusion from the \{111\} facets are the lowest among the inter-facets diffusion processes, indicating a more expressed tendency to expansion compared to \{100\} and \{110\} terminations. For the truncated cubooctahedron these differences are more pronounced than for the triangular cluster, bacause of the small initial dimensions of the \{111\}, which occupy the corners of the cluster.

\section{\label{sec:Conclusions}Conclusions}
In this work we assessed two known embedded atom model (EAM) parametrizations for the prediction of surfaces and Ag@Au nano-alloys properties. A particular interest was paid to the adsorption and diffusion processes, because of their fundamental role in the alloy growth, one of the main research topics for molecular dynamics and Monte Carlo simulations on nanoparticles. This validation was conduced on the parametrizations by Foiles and coworkers \cite{foiles_embedded-atom-method_1986} and by Zhou and coworkers,\cite{zhou_misfit-energy-increasing_2004} benchmarking the calculated properties with density functional theory calculations. \\
The two EAMs show small differences in the description of the processes on the surfaces, mainly concerning the characterization of the in-channel mechanisms, and overall agree with DFT calculations.\\
As expected, significant (and different depending on the parametrization) deviations were instead recovered in the small-scale. Here both EAMs fail in capturing some effects of the broken-periodicity in the adsorption energies, and sometimes do not predict possible diffusion paths, instead recovered from DFT calculations. Nevertheless, the two EAMs overall agree in the diffusion trends compared to the ones by DFT and can be considered to provide a faithful description of the interactions in these class of nano-systems.

\section{\label{sec:Acknowledgements} Acknowledgements}
The authors acknowledge funding from ERC under EU’s Horizon 2020 program (grant No. 681312). \\ 

\bibliography{aipsamp.bib}

\end{document}